\documentclass{article}

\usepackage{arxiv}          

\usepackage[utf8]{inputenc} 
\usepackage[T1]{fontenc}    
\usepackage{url}            
\usepackage{booktabs}       
\usepackage{amsfonts}       
\usepackage{nicefrac}       
\usepackage{microtype}      
\usepackage{lipsum}
\usepackage{float}
\usepackage{natbib}         
\usepackage{amsmath}        
\usepackage{fancyhdr}       
\usepackage{graphicx}       
\usepackage{siunitx}        
\usepackage{multirow}       
\usepackage{hyperref}       
\usepackage{graphicx}       
\usepackage{siunitx}        
\usepackage{cleveref}       

\graphicspath{{img/}}       

\title{Effective Benchmarks for Optical Turbulence Modeling} 

\author{ \href{https://orcid.org/0000-0003-0469-353X}{\includegraphics[scale=0.06]{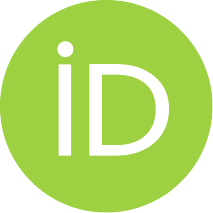}\hspace{1mm}Christopher ~Jellen}\thanks{Corresponding Author} \\
    Microsoft Corporation \\
	Redmond, WA 98052 \\
	\texttt{cjellen@microsoft.com} \\
	\And
	John ~Burkhardt, Cody ~Brownell \\
    Mechanical Engineering Department\\
	Untied States Naval Academy\\
	Annapolis, MD 21402 \\
    \And
	Charles ~Nelson \\
	Electrical and Computer Engineering Department\\
	Untied States Naval Academy\\
	Annapolis, MD 21402 \\
}
\date{January 3, 2024}

\renewcommand{\shorttitle}{Effective Benchmarks for Optical Turbulence}

\hypersetup{
pdftitle={Effective Benchmarks for Optical Turbulence Modeling},
pdfsubject={physics.ao-ph, cs.ai},
pdfauthor={Christopher ~Jellen, Charles ~Nelson, John ~Burkhardt, Cody ~Brownell},
pdfkeywords={benchmark, optical turbulence, boundary layer, scintillation, machine learning},
}

\begin{document} 
\maketitle
\setcitestyle{numbers}

\begin{abstract} 

Optical turbulence presents a significant challenge for communication, directed energy, and imaging systems, especially in the atmospheric boundary layer. Effective modeling of optical turbulence strength is critical for the development and deployment of these systems. The lack of standard evaluation tools, especially long-term data sets, modeling tasks, metrics, and baseline models, prevent effective comparisons between approaches and models. This reduces the ease of reproducing results and contributes to over-fitting on local micro-climates. Performance characterized using evaluation metrics provides some insight into the applicability of a model for predicting the strength of optical turbulence. However, these metrics are not sufficient for understanding the relative quality of a model. We introduce the \texttt{otbench} package, a Python package for rigorous development and evaluation of optical turbulence strength prediction models. The package provides a consistent interface for evaluating optical turbulence models on a variety of benchmark tasks and data sets. The \texttt{otbench} package includes a range of baseline models, including statistical, data-driven, and deep learning models, to provide a sense of relative model quality. \texttt{otbench} also provides support for adding new data sets, tasks, and evaluation metrics. The package is available at \url{https://github.com/cdjellen/otbench}. 

\end{abstract}

\section{Introduction} 

Optical turbulence is an important consideration in the design and deployment of communication, directed energy, and imaging systems \cite{andrews2005laser}. These systems must operate in a diverse set of propagation environments, and across a wide range of turbulent intensities \cite{andrews2005laser}. Effective modeling of optical turbulence strength in these propagation environments is a key requirement for operational effectiveness and reliability. Modeling optical turbulence in the boundary layer from macro-meteorology is exceptionally challenging; a wide range of models exist to address this problem \cite{app12104892} \cite{frederickson2000estimating} \cite{mahon2020comparison} \cite{sadot1992forecasting} \cite{wang2015prediction} \cite{wang2016using} \cite{jellen2020machine} \cite{jellen2023hybrid} \cite{vorontsov2020atmospheric}. While these models are often effective in their local environments during short-term measurement campaigns; understanding the applicability of these models across different and varied environments is an area of open research \cite{mahon2020comparison} \cite{jellen2023hybrid}. Additionally, robust evaluation of these models, as well as future models, is essential in operationalizing macro-meteorological models. 

The challenge of evaluating and comparing macro-meteorological models for optical turbulence intensity motivates the development of \texttt{otbench}. The package includes rich data sets from a range of field experiments, alongside well-defined modeling tasks and evaluation metrics. By coupling a data set, an objective, and an evaluation metric as a modeling task, \texttt{otbench} enables researchers to evaluate their models quickly and effectively against existing baselines, and to understand the performance of their approaches across a wider range of propagation environments. 

In addition to enabling consistent comparisons of macro-meteorological models for optical turbulence, \texttt{otbench} also includes the first publicly available long-term measurement campaign, collected over more than two years using the United States Naval Academy scintillometer link. This data set and the associated modeling tasks enable researchers to evaluate existing models for predicting and forecasting optical turbulence in the near-maritime boundary layer. 

The \texttt{otbench} package can be extended to include new data sets and baseline models. Each new data set and baseline model provides researchers with a deeper understanding of their model’s performance in context. By adopting \texttt{otbench}, researchers are empowered to evaluate their model across a diverse range of propagation environments and make rigorous comparisons against prior approaches. 

\section{Overview of \texttt{otbench}} 

Effective performance benchmarks require fixed objectives, evaluation metrics, and data sets. In the context of data-driven and deep learning models, this also includes fixed and transparent training, testing, and validation splits for each data set. Packages such as \texttt{WeatherBench} and the Open Graph Benchmark \texttt{ogb} provide researchers with a foundation for weather forecasting and for graph learning respectively \cite{rasp2020weatherbench} \cite{hu2020open}. Unfortunately, no such package exists for benchmarking optical turbulence modeling approaches. The \texttt{otbench} package works to bridge this gap. 

\subsection{Common tasks for modeling turbulence strength} 

A range of models exist to predict the strength of boundary layer optical turbulence from meteorological data \cite{sadot1992forecasting} \cite{wang2015prediction} \cite{wang2016using} \cite{pierzyna2023pi} \cite{frederickson2000estimating} \cite{jellen2023hybrid}. Establishing comparisons between these models requires careful re-implementation by each successive researcher. The lack of accessible field data further limits reproducibility. Open data, explicit model implementations, and reproducible predictions are essential in evaluating new models in context, and in making comparisons against prior approaches. 

The \texttt{otbench} package is built around \textit{tasks}, under which data sets are processed and standardized to enable comparisons across approaches. Data sets are developed from real-world field experiments and include a range of macro-meteorological features. Research teams currently decide how to re-sample, interpolate, or otherwise transform the data captured from these field experiments. This prevents effective comparisons between the models these teams develop. The \texttt{otbench} package uses tasks to enforces standard processing of data sets to mitigate this challenge. Tasks are explicit in their handling of missing measurements, data loading, and feature availability. Each task also references the model's target (measured $C_n^2$ at a given height of observation) and the evaluation metrics under which a model's performance is measured. This enables effective comparison between models, and understanding of a model's relative performance against prior literature models. 

\subsection{Domain diversity} 

The absence of standard benchmark tasks and data sets, especially long-term data sets, contribute to the challenge of over-fitting to local micro-climates. The \texttt{otbench} package attempts to address both of these concerns. The package currently includes data from both the Mona Loa $C_n^2$ Study and the United States Naval Academy Long Term Scintillation Study. Both data sets measure $C_n^2$ in the atmospheric boundary layer alongside local macro-meteorological data. Modeling approaches which perform well on both data sets may be robust to a wider range of propagation environments.

The United States Naval Academy Long Term Scintillation Study includes over two years of scintillometer data along with a range of macro-meteorological and oceanography parameters \cite{jellen2020measurement} \cite{Jellen:21}. This temporal range ensures models are evaluated across a range of seasonal conditions. In the context of machine learning and data-driven models, it also enables deeper investigation into the relationship between time and prediction error.

\subsection{Using \texttt{otbench}}

\begin{figure}[H]
    \centering
    \includegraphics[scale=0.48]{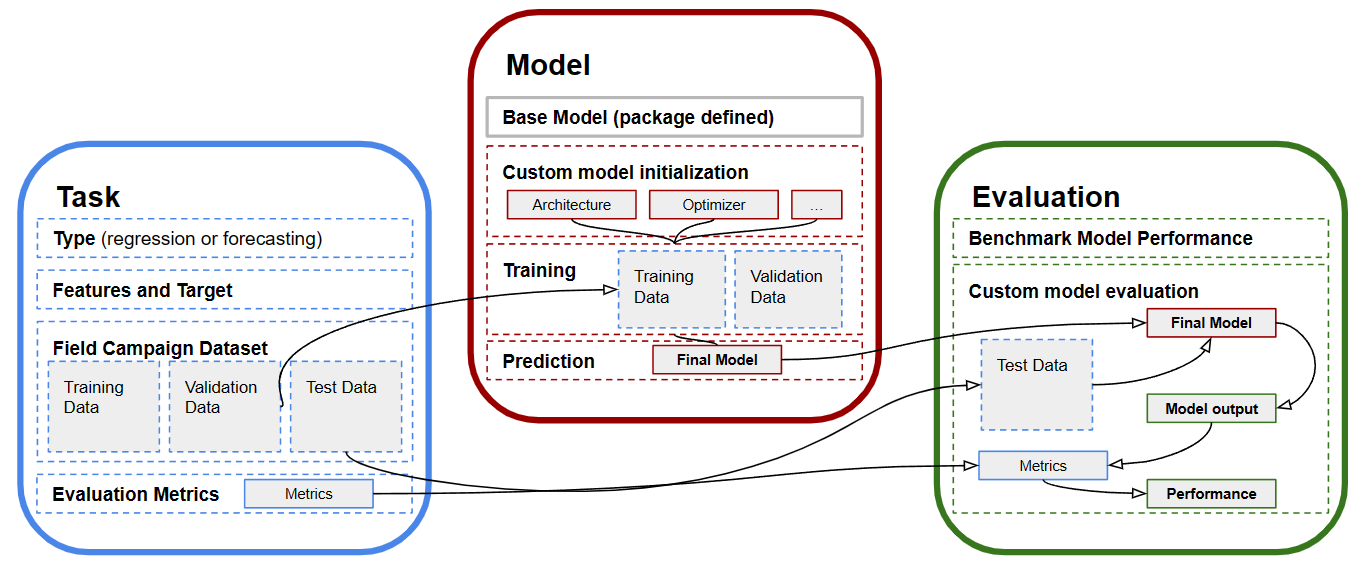}
    \caption{Model development and benchmark workflow under the \texttt{otbench} package.}
    \label{fig:otbench-diagram}
\end{figure}

The \texttt{otbench} package seeks to lower barriers to entry in evaluating existing models and developing new models for optical turbulence strength prediction. In addition to supplying benchmark tasks and data sets, \texttt{otbench} includes utilities for defining new models and evaluating their performance against a given task. Data-driven and deep learning models may be implemented as sub-classes of the base regression or forecasting model, offering data-processing and evaluation utilities provided by \texttt{otbench}. Macro-meteorological models with explicit parameters are also easy to develop under this context. The package includes sample implementations of each type of model, further described in \Cref{sec:baseline_models}. By adopting \texttt{otbench}, researchers are empowered to focus on crafting new models, with the package defining the data processing, model evaluation, and providing performance metrics for past implementations. This development flow is described visually in \Cref{fig:otbench-diagram}.

In \Cref{fig:otbench-diagram}, the "Task" and "Evaluation" phases are defined by the \texttt{otbench} package, while the "Model" phase is defined by the researcher, optionally using the package-defined "Base model" for macro-meteorological, data-driven, or deep learning approaches. This base model supplies standard utilities to increase the velocity of model development. For deep learning models, researchers need only specify a model architecture and \texttt{train} method; the base model supplies data loading functionality. Physics-based approaches which do not rely on local data for training can be defined as a "final model" without the need for other user-defined components. The "Evaluation" phase can evaluate model performance in both cases, optionally storing performance results for contextualization and comparison. For all user-defined models, the "Evaluation" is performed against the task's test data. This imposes a requirement that the user-defined model generate predictions without the use of atmospheric data not measured during the underlying field study.

\section{Data} 

\subsection{Mona Loa Observatory (MLO) $C_n^2$ Study} 

Optical turbulence is a key concern for earth-based observatories. The High Altitude Observatory (HAO) manages a solar observatory at Mona Loa, Hawai’i \cite{10-26023-CQR2-TQJ9-AH10} \cite{macleanoncley2006}. The HAO and the National Center for Atmospheric Research (NCAR) conducted a measurement campaign to study optical turbulence in the boundary layer during the summer of 2006 with the support of the Integrated Surface Flux Facility (ISFF) \cite{10-26023-CQR2-TQJ9-AH10}. This study used sonic anemometers fixed at a series of heights to quantify the refractive index structure parameter $C_n^2$, along with meteorological features. Instruments were deployed on an existing 40\si{\metre} tower operated by the National Oceanographic and Atmospheric Administration (NOAA) at the Mauna Loa Observatory. Sonic anemometers were fixed on the east side of the tower at heights of 6.045\si{\metre}, 13.3\si{\metre}, and 23.2\si{\metre} from the base of the tower \cite{10-26023-CQR2-TQJ9-AH10}. The sonic's data was supplemented by additional NOAA data which measured atmospheric features near the surface, with wind data near 10\si{\metre}. Data from this measurement campaign is made available through the University Corporation for Atmospheric Research (UCAR) at \url{https://data.eol.ucar.edu/data set/160.007}. 

The data is distributed under the NetCDF4 file format \cite{10-5065-D6H70CW6}. Measurements are averaged at a 5 minute frequency, and were collected between June 9\textsuperscript{th}, 2006 and August 8\textsuperscript{th}, 2006. An overview of the available data, including the heights of measurement on the observation tower, are available in \Cref{table:1}.

\begin{table}[H]
    \centering
    \caption{Overview of the MLO $C_n^2$ data set.}
    \begin{tabular}{|| m{4cm} || >{\centering}m{1cm} | m{5cm} | >{\centering}m{1.2cm} | >{\centering\arraybackslash}m{2cm} ||}
        \hline
        \textbf{Long Name} & \textbf{Short Name} & \textbf{Description} & \textbf{Units} & \textbf{Heights [\si{\metre}]} \\
        \hline
        \hline
        Raw voltage & kh2oV & Raw voltage & \si{\volt} & 6, 15, 25 \\
        Water vapor density & kh2o & Water vapor density & \si{\gram\per\cubic\metre} & 6, 15, 25 \\
        U component & u & U component of wind speed & \si{\metre\per\second} & 6, 15, 25 \\
        V component & v & V component of wind speed & \si{\metre\per\second} & 6, 15, 25 \\
        W component & w & W component of wind speed & \si{\metre\per\second} & 6, 15, 25 \\
        Virtual temperature & tc & Virtual temperature from the speed of sound & \si{\celsius} & 6, 15, 25 \\
        Wind direction & Dir & Mean wind direction & \si{\celsius} & 10 \\
        Wind speed & Spd & Mean wind speed & \si{\metre\per\second} & 10 \\
        Air pressure & P & Barometric pressure & \si{\milli\bar} & 2 \\
        Air temperature & T & Ambient air temperature & \si{\celsius} & 2 \\
        Relative humidity & RH & Relative humidity & \si{\%} & 2 \\
        Dew point temperature & Tdew & Dew point temperature & \si{\celsius} & 2 \\
        Refractive index structure parameter & $C_n^2$ & Optical turbulence strength & \si{\metre\textsuperscript{$-\frac{2}{3}$}} & 6, 15, 25 \\
         \hline
         \multicolumn{5}{|c|}{\textbf{Coordinates}} \\ 
         \hline
         Time & time & The local time at which a measurement was taken & \si{\second} & - \\
        \hline
    \end{tabular}
    \label{table:1}
\end{table}

In \Cref{table:1}, 9,857 of the 14,038 total rows contain no missing measurements. 

\subsection{Unites States Naval Academy long-term scintillation study} 

The USNA long term scintillation study is a continuing effort to characterize and measure optical turbulence in the near-maritime boundary layer. A range of experiments have been conducted across the Severn River in Annapolis, Maryland \cite{Jellen:21}. The longest running of these experiments used a ScinTec BLS450 scintillometer to characterize the strength of optical turbulence over a link approximately 3\si{\metre} above the river’s surface between January 1\textsuperscript{st}, 2020, and July 26\textsuperscript{th}, 2023 \cite{jellen2023hybrid} \cite{scintecbls450}. This field campaign included a range of meteorological and oceanographic measurements from nearby NOAA, National Data Buoy Center (NDBC), and Academy weather stations.

The two subsets of this campaign are used as data sets. The first, denoted \texttt{usna\_cn2\_sm}, includes 3 months in the summer of 2021 in which few measurements were missing from the scintillometer data set, supplemented with NOAA Coastal Observation (COOPS) data describing the local atmosphere and water conditions \cite{noaacoops85755122023}. The second covered the period between January 1\textsuperscript{st}, 2020 and July 14\textsuperscript{th}, 2022 and is denoted \texttt{usna\_cn2\_lg}. This data set includes local measurements made with a Davis Vantage Pro2 weather station, as well as water conditions from the NDBC Thomas Point data buoy \cite{davisvantagepro22023} \cite{noaandbctplm22023}. Future releases will develop a full 42 month data set, including measurements from all sources as available. 

\subsubsection{USNA $C_n^2$ short-duration data set} 

Field campaigns studying optical turbulence present a significant expense for researchers. Some capture measurements over a small number of days or weeks \cite{sadot1992forecasting} \cite{wang2015prediction} using bulk atmospheric weather stations rather than high-frequency sonics. The low cost and availability of these weather stations may present an opportunity to develop models which perform across disparate propagation environments. 

The \texttt{usna\_cn2\_sm} data set includes measured $C_n^2$ along with atmospheric and oceanographic parameters from a co-located NOAA COOPS observation station between June 1\textsuperscript{st}, 2021, and September 1\textsuperscript{st}, 2021. These summer months were selected to roughly map to the time-of-year used in the MLO $C_n^2$ study, and due to the low number of missing measurements from the scintillometer. The \texttt{usna\_cn2\_sm} data set includes parameters at a single level of observation, with a 6 minute frequency of observation. A detailed description is presented in \Cref{table:2}. 

\begin{table}[H]
    \centering
    \caption{Overview of the USNA $C_n^2$ Small data set.}
    \begin{tabular}{|| m{4cm} || >{\centering}m{1cm} | m{5cm} | >{\centering}m{1.2cm} | >{\centering\arraybackslash}m{2cm} ||}
        \hline
        \textbf{Long Name} & \textbf{Short Name} & \textbf{Description} & \textbf{Units} & \textbf{Heights [\si{\metre}]} \\
        \hline
        \hline
        Wind direction & Dir & Mean wind direction & \si{\metre\per\second} & 10 \\
        Wind speed & Spd & Mean wind speed & \si{\metre\per\second} & 10 \\
        Air pressure & P & Barometric pressure & \si{\milli\bar} & 10 \\
        Air temperature & T & Ambient air temperature & \si{\celsius} & 5 \\
        Relative humidity & RH & Ambient relative humidity & \si{\%} & 2 \\
        Water temperature & T & Water surface temperature & \si{\celsius} & 0 \\
        Solar radiation & Rad & Incoming total solar radiation at ground level & \si{\watt\per\squared\metre} & 1 \\
        Refractive index structure parameter & $C_n^2$ & Optical turbulence strength & \si{\metre\textsuperscript{$-\frac{2}{3}$}} & 3 \\
         \hline
         \multicolumn{5}{|c|}{\textbf{Coordinates}} \\ 
         \hline
         Time & time & The local time at which a measurement was taken & \si{\second} & - \\
         Latitude & lat & The latitude at which the scintillometer and weather station are located & \si{\degree} & - \\
         Longitude & lon & The longitude at which the scintillometer and weather station are located & \si{\degree} & - \\
         Altitude & alt & The height above sea level of the scintillometer & \si{\metre} & - \\
        \hline
    \end{tabular}
    \label{table:2}
\end{table}

In \Cref{table:2}, 21,007 of the 22,081 total observations have no missing measurements. As a result, no interpolation is applied to the data set. The \texttt{usna\_cn2\_sm} data set captures the rapid fluctuations in $C_n^2$, as well as the weaker diurnal cycle for optical turbulence in the near-maritime environment above the Severn River. 

\subsubsection{USNA $C_n^2$ long-duration data set} 

In contrast to the \texttt{usna\_cn2\_sm} data set, the \texttt{usna\_cn2\_lg} data set includes data captured or interpolated to a 1 minute frequency, with some significant gaps due to instrument misalignment or power outages. This data set offers the first known opportunity to develop models for predicting or forecasting optical turbulence over more than two years. The periods of missing measurements present an additional challenge for model development, better reflecting real-world conditions. A detailed description is presented in \Cref{table:3}.

\begin{table}[H]
    \centering
    \caption{Overview of the USNA $C_n^2$ Large data set.}
    \begin{tabular}{|| m{4cm} || >{\centering}m{1cm} | m{5cm} | >{\centering}m{1.2cm} | >{\centering\arraybackslash}m{2cm} ||}
        \hline
        \textbf{Long Name} & \textbf{Short Name} & \textbf{Description} & \textbf{Units} & \textbf{Heights [\si{\metre}]} \\
        \hline
        \hline
        Wind direction & Dir & Mean wind direction & \si{\metre\per\second} & 3 \\
        Wind speed & Spd & Mean wind speed & \si{\metre\per\second} & 3 \\
        Air pressure & P & Barometric pressure & \si{\milli\bar} & 3 \\
        Air temperature & T & Ambient air temperature & \si{\celsius} & 3 \\
        Relative humidity & RH & Ambient relative humidity & \si{\%} & 3 \\
        Water temperature & T & Water surface temperature & \si{\celsius} & 0 \\
        Solar radiation & Rad & Incoming total solar radiation at ground level & \si{\watt\per\squared\metre} & 1 \\
        Temporal hour & th & The time elapsed since sunrise divided by 1/12 the time between sunset and sunrise for a given day & - & - \\
        Refractive index structure parameter & $C_n^2$ & Optical turbulence strength & \si{\metre\textsuperscript{$-\frac{2}{3}$}} & 3 \\
         \hline
         \multicolumn{5}{|c|}{\textbf{Coordinates}} \\ 
         \hline
         Time & time & The local time at which a measurement was taken & \si{\second} & - \\
         Latitude & lat & The latitude at which the scintillometer and weather station are located & \si{\degree} & - \\
         Longitude & lon & The longitude at which the scintillometer and weather station are located & \si{\degree} & - \\
         Altitude & alt & The height above sea level of the scintillometer & \si{\metre} & - \\
        \hline
    \end{tabular}
    \label{table:3}
\end{table}

In \Cref{table:3}, 1,155,041 of the 1,291,225 total observations have no missing measurements after interpolation to re-index to a 1 minute frequency. This data set presents an opportunity to study model performance across seasons and over longer time scales. It further presents an opportunity to assess a model over a validation set of one year or longer, while training with data across all seasons. 

\subsection{Adding new field campaigns}

As currently implemented, \texttt{otbench} includes three data sets from two field campaigns. While this provides researchers with a measure of environmental extensibility for their models, new data sets captured in different propagation environments will increase validation quality. The package currently stores data sets under the NetCDF4 format, using time as a required coordinate. There are no specific requirements for the instruments used to capture data or handling of missing measurements. A standard frequency of observation is assumed when developing and evaluating forecasting models; however, the package does not impose strict requirements on formatting or metadata requirements on its constituent data sets. New data sets are best contributed directly to the source repository at \url{https://github.com/cdjellen/otbench}. 

\section{Benchmark Tasks} 

While standard data sets are necessary for comparing model performance, especially across the literature, they are not sufficient for robust evaluation. Fixed training and validation sets for data-driven modeling approaches, a fixed test set for all models, standard evaluation metrics, and standard handling of missing measurements improve the quality and robustness of benchmark evaluation. The \texttt{otbench} package includes regression tasks and forecasting tasks. Under regression tasks, models predict the strength of optical turbulence from a set of macro-meteorological measurements. In forecasting tasks, the measured $C_n^2$ some fixed number of observations in the future is predicted from past macro-meteorological and $C_n^2$ measurements. 

All tasks have common structure, specifying a data set, metadata, treatment of missing measurements (either \texttt{“full”} for the complete data set or \texttt{“dropna”} for the data set without missing measurements). These observations are removed from the train, test, and validation sets independently. Each task further specifies the target parameter, typically the name of the $C_n^2$ column for that data set (such as \texttt{“Cn2\_3m”} or \texttt{“Cn2\_15m”}), and the columns assumed unavailable for training purposes, typically $C_n^2$ measured at other elevations. Some models which make use of solar features such as the temporal hour require knowledge of the location and timezone of the data set; these are specified at the task level. Finally, tasks specify the train, test, and validation sets based on the index of the transformed data set, whether to take the base-10 log of $C_n^2$, and which metrics to evaluate model performance under. 

\subsection{Regression}
\label{sec:benchmark_tasks__regression}

Regression tasks use a set of standard metrics including the Root Mean Square Error (RMSE), Coefficient of Determination ($R^2$), Mean Absolute Error (MAE), and Mean Absolute Percentage Error (MAPE).  These metrics were selected in an effort to enable comparisons against prior literature \cite{pierzyna2023pi}. For a given regression task, all models are compared across these metrics and the number of valid (non-null) predictions generated by the model. This context offers researchers insight into the relative performance of existing models, as well as the potential improvement afforded by the new models which their teams develop. The current regression tasks are further described in \Cref{table:4}. 

\begin{table}[H]
    \centering
    \caption{Current regression tasks implemented for \texttt{otbench}.}
    \begin{tabular}{|| m{2cm} || >{\centering}m{3cm} | >{\centering}m{3cm} | >{\centering}m{3cm} | >{\centering\arraybackslash}m{3cm} ||}
        \hline
         & \textbf{MLO $C_n^2$ (full)} & \textbf{MLO $C_n^2$ (dropna)} & \textbf{USNA $C_n^2$ Small} & \textbf{USNA $C_n^2$ Large} \\
        \hline
        \hline
        Data set name & \texttt{mlo\_cn2} & \texttt{mlo\_cn2} & \texttt{usna\_cn2\_sm} & \texttt{usna\_cn2\_lg} \\
        Latitude & 19.53 & 19.53 & 38.98 & 38.98 \\
        Longitude & -155.57 & -155.57 & -176.48 & -176.48 \\
        Time zone & US/Hawaii & US/Hawaii & US/Eastern & US/Eastern \\
        Train Indices & \texttt{[0, 8366]} & \texttt{[0, 8366]} & \texttt{[0, 14639]} & \texttt{[0, 524339]} \\
        Test Indices & \texttt{[8367, 10366]} & \texttt{[8367, 10366]} & \texttt{[14640, 17999]} & \texttt{[524340, 788208]} \\
        Validation Indices & \texttt{[10367, 13942]} & \texttt{[10367, 13942]} & \texttt{[18000, 22080]} & \texttt{[788209, 1291224]} \\
        Drop missing & \texttt{false} & \texttt{true} & \texttt{false} & \texttt{false} \\
        Log transform & \texttt{true} & \texttt{true} & \texttt{true} & \texttt{true} \\
        Target & $C_n^2$ \si{15 \metre} & $C_n^2$ \si{15 \metre} & $C_n^2$ \si{3 \metre} & $C_n^2$ \si{3 \metre} \\
        \hline
    \end{tabular}
    \label{table:4}
\end{table}

\subsection{Forecasting} 

Forecasting tasks differ from regression tasks by including a fixed inference window size $m$ and forecast horizon $n$. The window size specifies the number of past observations available at inference time to predict the strength of optical turbulence $n$ observations in the future. This window includes the past $m$ observations of both macro-meteorological parameters and $C_n^2$, in contrast to regression tasks which do not include $C_n^2$ at inference time. 

Forecasting tasks assume direct forecasting, in which the extent of $C_n^2$ $n$ observations in the future is predicted from available data. This is in contrast to iterative forecasting, in which $C_n^2$ is predicted 1 observation in the future, and then fed back into the model until the predicted extent of $C_n^2$ at $n$ is available. The current forecasting tasks are further described in \Cref{table:5}. Forecasting tasks implement the same evaluation metrics as regression tasks described in \Cref{sec:benchmark_tasks__regression}.

\begin{table}[H]
    \centering
    \caption{Current forecasting tasks implemented for \texttt{otbench}.}
    \begin{tabular}{|| m{2cm} || >{\centering}m{4cm} | >{\centering\arraybackslash}m{4cm} ||}
        \hline
         & \textbf{MLO $C_n^2$ (dropna)} & \textbf{USNA $C_n^2$ Small} \\
        \hline
        \hline
        Data set name &  \texttt{mlo\_cn2} & \texttt{usna\_cn2\_sm} \\
        Latitude & 19.53 & 38.98 \\
        Longitude & -155.57 & -176.48 \\
        Time zone & US/Hawaii & US/Eastern \\
        Train Indices  & \texttt{[0, 8366]} & \texttt{[0, 14639]}\\
        Test Indices  & \texttt{[8367, 10366]} & \texttt{[14640, 17999]} \\
        Validation Indices & \texttt{[10367, 13942]} & \texttt{[18000, 22080]} \\
        Drop missing & \texttt{true} & \texttt{true} \\
        Log transform & \texttt{true} & \texttt{true} \\
        Target & $C_n^2$ \si{15 \metre} & $C_n^2$ \si{3 \metre} \\
        \hline
    \end{tabular}
    \label{table:5}
\end{table}

\section{Baseline models}
\label{sec:baseline_models} 

The package includes a range of literature and data-driven models as baselines for assessing the relative quality of new models. These models fall under three broad categories; models inspired by weather forecasting including climatology and persistence \cite{rasp2020weatherbench}, prior macro-meteorological models developed from in-situ measurements \cite{sadot1992forecasting} \cite{wang2015prediction} \cite{chen2019climatological} \cite{jellen2023hybrid}, and simple deep learning models.

Regression models and forecasting models are implemented separately, enabling forecasting models to leverage prior observations of the target to inform predictions at the forecast horizon. These prior measurements are typically unavailable for operational regression models, which predict optical turbulence strength from other measurable parameters. The set of baseline models can be extended to cover more of the literature and novel deep learning approaches to the optical turbulence regression and forecasting tasks. Although not required for future tasks, all current tasks transform measured $C_n^2$ to the corresponding $\log{10} C_n^2$. 

\subsection{Statistical models}

Baseline models provide the context required to characterize the performance of more complex models. The \texttt{WeatherBench} package implements persistence and climatology approaches for the task of weather forecasting \cite{rasp2020weatherbench}. These inspire the persistence and climatology models in \texttt{otbench}. In the context of forecasting, these baselines are augmented by simple linear forecasting and mean-window models. The linear forecasting model uses least-squares regression to fit a curve to the window of data available during inference, while the mean window model predicts the mean value of $C_n^2$ observed during this window.

\subsubsection{Persistence}

Persistence models serve as a common baseline in weather forecasting contexts \cite{rasp2020weatherbench}. These models predict a static value for the target, in this case $C_n^2$, for all future observations. In the context of optical turbulence modeling, persistence models take the most recent observation of $C_n^2$ as the prediction for the next observation. Other models are expected to outperform persistence models, as they do not leverage any information about the macro-meteorological conditions at the time of prediction. Given the diversity of propagation environments, persistence models provide useful context for understanding the performance of new and existing optical turbulence models.

\subsubsection{Climatology}

Rather than predicting the most recent observation of $C_n^2$ as in the case of persistence models, climatology models compute a relevant mean $C_n^2$ value from the training set and use this as the prediction for all future observations. In the context of optical turbulence modeling, climatology models compute the mean $C_n^2$ for all observations in the training set, excluding missing measurements, and use this as the prediction for all future observations. The \texttt{otbench} package implements both a standard climatology model as well as a climatology model which computes a mean value for each minute of the day. The minute climatology model uses the mean value for the given clock time seen in training as its prediction at inference time. Climatology models are expected to outperform persistence models. Their inclusion as benchmarks in the \texttt{otbench} package provides a sense of the relative quality of new and existing models, especially in the context of specific propagation environments.

\subsection{Data-driven models}

Measurement campaigns enable researchers to fit optical turbulence models from data. These macro-meteorological models are a common approach to predicting the strength of optical turbulence from bulk measurements \cite{sadot1992forecasting} \cite{wang2015prediction} \cite{chen2019climatological}. These models are fit using data acquired from field campaigns, and are capable of generating point predictions for $C_n^2$ for a given measurement set. Macro-meteorological approaches present two key challenges; they implicitly define a set of required meteorological parameters and they present a risk of over-fitting to the local micro-climate \cite{wang2015prediction} \cite{jellen2020measurement}. Some models, such as those in \cite{sadot1992forecasting} and \cite{wang2015prediction}, are parametric, while those in \cite{jellen2020measurement} and \cite{jellen2023hybrid} are non-parametric. The Gradient Boosting Regression Tree (GBRT) architecture presented in \cite{ke2017lightgbm} and the Hybrid air-water temperature difference model presented in \cite{jellen2023hybrid} provide baseline metrics for data-driven, tree-based models as applied to the task of optical turbulence strength prediction and forecasting. 

\subsubsection{The macro-meteorological model}

The authors of \cite{sadot1992forecasting} fit a model for $C_n^2$ in an over-land propagation environment at a height of 15\si{\metre}:

\begin{equation}
    \begin{aligned}
        C_n^2 & = (3.8 \times 10^{-14})W + f(T) + f(U) + f(RH) -(5.3 \times 10^{-13}) \\
        \text{where} \\
        f(T) & = (2.0 \times 10^{-15})T \\
        f(U) & = ( -2.5 \times 10^{-15})U + (1.2 \times 10^{-15})U^{2} -(8.5 \times 10^{-15})U^{3} \\
        f(RH) & = ( -2.8 \times 10^{-15})RH + (2.9 \times 10^{-17})RH^{2} -(1.1 \times 10^{-19})RH^{3} \\
    \end{aligned}
\end{equation}
\label{eq1}

In \Cref{eq1}, $W$ denotes the temporal hour weight \cite{sadot1992forecasting}, $T$  denotes the temperature in \si{\kelvin}, $RH$ denotes the relative humidity in \si{\%}, and $U$ denotes the wind speed in \si{\metre\per\second}. In order to generate predictions from this model, the dynamic range of observation presented in \cite{sadot1992forecasting} was enforced. Any measurement in which any meteorological parameters were outside of the dynamic range was dropped from the training and validation sets.

\subsubsection{The offshore macro-meteorological model}

The macro-meteorological model in \cite{sadot1992forecasting} was evaluated by the authors of \cite{wang2015prediction} in a coastal environment. The authors additionally fit a model for $C_n^2$ using data from their measurement campaign, reproduced as:

\begin{equation}
    \begin{aligned}
        C_n^2 & = (-1.58 \times 10^{-15})W + f(T) + f(U) + f(RH) -(7.44 \times 10^{-14}) \\
        \text{where} \\
        f(T) & = (2.74 \times 10^{-16})T \\
        f(U) & = (3.37 \times 10^{-16})U + (1.92 \times 10^{-16})U^{2} -(2.8 \times 10^{-17})U^{3} \\
        f(RH) & = (8.3 \times 10^{-17})RH - (2.22 \times 10^{-18})RH^{2} + (1.42 \times 10^{-20})RH^{3} \\
    \end{aligned}
\end{equation}
\label{eq2}

In \Cref{eq2}, $W$ denotes the temporal hour weight \cite{sadot1992forecasting} $T$ denotes the temperature in \si{\kelvin}, $RH$ denotes the relative humidity in \si{\%}, and $U$ denotes the wind speed in \si{\metre\per\second}. As for the macro-meteorological model in equation (\ref{eq1}), the dynamic range of the measurement campaign in \cite{wang2015prediction} was enforced when generating model predictions.

\subsubsection{The air-water temperature difference model}

The impact of the air-water temperature difference on the strength of local optical turbulence was noted in \cite{frederickson2000estimating}, \cite{jellen2020measurement}, and \cite{chen2019climatological}. The authors of \cite{chen2019climatological} reproduced a model for predicting $C_n^2$ at ground level as a function of only the measured air-water temperature difference in \si{\celsius}:

\begin{equation}
    C_n^2(0) = ({2.05\Delta T}^2 + 2.37\Delta T + 1.58) \times 10^{-16}
\end{equation}
\label{eq3}

In \Cref{eq3}, all observations in which both the air temperature and water temperature were available result in a prediction for $C_n^2$. These predictions were scaled to the task-appropriate height using the approach described in \cite{jellen2023hybrid}.

\subsection{Deep learning models}

Deep learning models are an area of active research in the field of optical turbulence modeling \cite{vorontsov2020atmospheric} \cite{wang2016using}. These models may offer improved performance of prior approaches, especially in contexts where high-frequency or long-term data is available for training \cite{vorontsov2020atmospheric}. The potential applicability of deep learning models for optical turbulence strength modeling motivates their inclusion as baselines in \texttt{otbench}.
The \texttt{otbench} package includes a set of utilities to enable deep learning model development in forecasting and regression tasks. These utilities are built using the \texttt{PyTorch} framework \cite{paszke2019pytorch}. Through the inclusion of these tools, \texttt{otbench} serves as a starting point for researchers interested in developing novel deep learning models.

\subsubsection{The recurrent neural network model}

A basic, single-module recurrent neural network (RNN) architecture \cite{paszke2019pytorch}, based on the implementation described in \cite{elman1990finding}, is available for both regression and forecasting across all tasks. Optical turbulence data sets include sequential measurements of $C_n^2$ and other relevant meteorological parameters. This architecture has been applied for other time-series modeling tasks \cite{elman1990finding}. While deeper models with effective hyper-parameter tuning may outperform the basic RNN model in \texttt{otbench}, this model presents a minimum baseline for comparison against future approaches.

\section{Baseline model performance}

\subsection{Regression}

Existing macro-meteorological models often focus on the challenge of predicting $C_n^2$ from bulk measurements \cite{sadot1992forecasting} \cite{wang2015prediction} \cite{wang2016using} \cite{chen2019climatological}. These models are fit using data acquired from field campaigns and are capable of generating point predictions for $C_n^2$ for a given measurement set. Under the \texttt{otbench} package, these are implemented as regression, and require the same set of macro-meteorological parameters used when the model was initially fit. Some of these models leverage solar features, such as the temporal hour or temporal hour weight, derived from the local time of observation and the local sunrise time \cite{sadot1992forecasting} \cite{wang2015prediction} \cite{jellen2020measurement} \cite{jellen2023hybrid}.

\subsubsection{\texttt{mlo\_cn2}}
\label{sec:regression_mlo_cn2} 

The two regression tasks implemented using the \texttt{mlo\_cn2} data set include benchmarks for persistence, climatology, macro-meteorological, and data driven models. As the atmospheric boundary layer was not over water, the models in \cite{chen2019climatological} and \cite{jellen2023hybrid} are not included as benchmarks for these tasks. The performance of each benchmark model is described in \Cref{table:6} for the task in which missing measurements are removed, and \Cref{table:7} for the task in which they are not. 

\begin{table}[H]
    \centering
    \caption{Benchmark model performance for the \texttt{mlo\_cn2} regression task, removing missing measurements.}
    \begin{tabular}{|| m{4.5cm} || >{\centering}m{1.2cm} | >{\centering}m{1.2cm} | >{\centering}m{1.2cm} | >{\centering}m{1.5cm} | >{\centering\arraybackslash}m{3.5cm} ||}
        \hline
        \textbf{Model} & \textbf{RMSE} & \textbf{MAE}& \textbf{MAPE} & \textbf{$R^2$} & \textbf{Valid predictions\newline(total possible)} \\
        \hline
        \hline
            Macro Meteorological          & 0.714 & 0.587 & 0.044 & 0.021 & 813 (2449) \\
            Offshore Macro Meteorological & 0.816 & 0.622 & 0.044 & 0.040 & 2274 (2449) \\
            Persistence       & 1.209 & 1.016 & 0.072 & 0.0 & 2449 (2449) \\
            Minute Climatology & 0.504 & 0.384 & 0.028 & 0.538  & 2449 (2449) \\
            Climatology       & 0.661 & 0.531 & 0.038 & 0.0 & 2449 (2449) \\
            GBRT  & 0.212 & 0.154 & 0.011 & 0.910  & 2449 (2449) \\ 
            RNN  & 0.336 & 0.213 & 0.015 & 0.761  & 2449 (2449) \\ 
        \hline
    \end{tabular}
    \label{table:6}
\end{table}

The persistence model is a common baseline for weather forecasting evaluation \cite{rasp2020weatherbench}. In \Cref{table:6}, the model uses the most recent observation, the final observation in the test set, as it’s prediction of $C_n^2$ at \si{15\metre}. All other benchmark models outperform this approach. The climatology model uses all observations in the training and test sets to compute the mean value of $C_n^2$ at \si{15\metre}, excluding missing measurements. This value is then used as the prediction for all subsequent observations. The climatology model outperforms the two macro-meteorological models, indicating that they may be over-fit to the propagation environments in which they were developed, or that they may be inapplicable for the conditions at Mona Loa during the study. This finding highlights the value of \texttt{otbench} as a tool for model evaluation and comparison; differential performance for a given modeling approach on a regression task requires outperforming the persistence and climatology models. The RNN model performed adequately in the regression task, demonstrating improvement over all persistence and climatology baselines. Additional architectural improvements and wider input windows may further improve the model's performance, however, the results in \Cref{table:6} provide a minimum baseline for deep learning models as applied to this MLO $C_n^2$ regression task. The random forest model used as a baseline in \cite{jellen2023hybrid} outperforms both the RNN and climatology models, demonstrating substantial prediction accuracy with an $R^2$ of 0.897, indicating a high level of explained variance in measured $C_n^2$ at \si{15\metre}. These results, with the exception of the performance of the un-tuned, baseline GBRT model, hold for the task in which missing measurements are not removed from the data set in \Cref{table:7}. 

\begin{table}[H]
    \centering
    \caption{Benchmark model performance for the \texttt{mlo\_cn2} regression task, full.}
    \begin{tabular}{|| m{4.5cm} || >{\centering}m{1.2cm} | >{\centering}m{1.2cm} | >{\centering}m{1.2cm} | >{\centering}m{1.5cm} | >{\centering\arraybackslash}m{3.5cm} ||}
        \hline
        \textbf{Model} & \textbf{RMSE} & \textbf{MAE}& \textbf{MAPE} & \textbf{$R^2$} & \textbf{Valid predictions\newline(total possible)} \\
        \hline
        \hline
            Macro Meteorological          & 0.620 & 0.514 & 0.039 & 0.032 & 862 (2816) \\
            Offshore Macro Meteorological  & 0.830 & 0.632 & 0.044 & 0.024 & 2641 (2816) \\
            Persistence & 1.221 & 1.030 & 0.073 & 0.0 & 2816 (2816) \\
            Minute Climatology  & 0.586 & 0.441 & 0.032 & 0.427 & 2816 (2816) \\
            Climatology     & 0.665 & 0.530 & 0.038 & 0.0 & 2816 (2816) \\
            GBRT & 4.503 & 3.147 & 0.230 & 0.065 & 2816 (2816)  \\
            RNN & 0.548 & 0.413 & 0.030 & 0.343 & 2816 (2816)  \\
        \hline
    \end{tabular}
    \label{table:7}
\end{table}

In \Cref{table:7}, the RNN model again outperformed the baseline persistence and climatology models, despite the presence of missing measurements in the data. The GBRT model performs poorly in the case in which missing measurements are present across observations in the training and test sets. The dramatic drop in model performance could result from the lack of adjustment to address patterns in missingness observed in the training data. 

\subsubsection{\texttt{usna\_cn2\_sm}} 

The scintillometer link at the United States Naval Academy is located across the Severn River, with well over 95\% of the total link length over water \cite{Jellen:21}. The water temperature is available as a feature, enabling two additional models over those in \ref{sec:regression_mlo_cn2}. The air-water temperature difference model is referenced in \cite{chen2019climatological}, while the hybrid model is introduced in \cite{jellen2023hybrid}. All benchmark regression models are evaluated in \Cref{table:8}. 

\begin{table}[H]
    \centering
    \caption{Benchmark model performance for the \texttt{usna\_cn2\_sm} regression task.}
    \begin{tabular}{|| m{4.5cm} || >{\centering}m{1.2cm} | >{\centering}m{1.2cm} | >{\centering}m{1.2cm} | >{\centering}m{1.5cm} | >{\centering\arraybackslash}m{3.5cm} ||}
        \hline
        \textbf{Model} & \textbf{RMSE} & \textbf{MAE}& \textbf{MAPE} & \textbf{$R^2$} & \textbf{Valid predictions\newline(total possible)} \\
        \hline
        \hline
            Air-water Temperature Difference & 0.910 & 0.788 & 0.056 & 0.267 & 4080 (4080) \\
            Macro Meteorological & 0.864 & 0.685 & 0.047 & 0.268 & 3653 (4080) \\
            Offshore Macro Meteorological & 0.178 & 0.126 & 0.009 & 0.058 & 421 (4080) \\
            Persistence & 0.758 & 0.682 & 0.049 & 0.0 & 4080 (4080) \\
            Minute Climatology & 0.452 & 0.362 & 0.026 & 0.126 & 4080 (4080) \\
            Climatology & 0.480 & 0.382 & 0.027 & 0.0 & 4080 (4080) \\
            Hybrid AWT & 0.303 & 0.218 & 0.015 & 0.583 & 4080 (4080) \\
            GBRT & 0.299 & 0.220 & 0.015 & 0.584 & 4080 (4080) \\
            RNN & 0.375 & 0.281 & 0.020 & 0.451 & 4080 (4080) \\
        \hline
    \end{tabular}
    \label{table:8}
\end{table}

Under the \texttt{usna\_cn2\_sm} regression task, the Offshore Macro-meteorological model presented in \cite{wang2015prediction}, the random forest baseline from \cite{jellen2020measurement} and \cite{jellen2020machine}, the RNN model built using \cite{paszke2019pytorch}, and the hybrid model introduced in \cite{jellen2023hybrid} all outperformed the persistence and climatology baselines. When compared to the results in \Cref{table:7}, the minute-level climatology model in \Cref{table:8} showed much lower performance improvement over the standard climatology model. 

\subsubsection{\texttt{usna\_cn2\_lg}} 

The \texttt{usna\_cn2\_lg} data set enables long-term performance evaluation for regression models. All models applicable to the regression task for the \texttt{usna\_cn2\_sm} data set also apply for the regression task built from the large data set. 

\begin{table}[H]
    \centering
    \caption{Benchmark model performance for the \texttt{usna\_cn2\_lg} regression task.}
    \begin{tabular}{|| m{4.5cm} || >{\centering}m{1.2cm} | >{\centering}m{1.2cm} | >{\centering}m{1.2cm} | >{\centering}m{1.5cm} | >{\centering\arraybackslash}m{3.5cm} ||}
        \hline
        \textbf{Model} & \textbf{RMSE} & \textbf{MAE}& \textbf{MAPE} & \textbf{$R^2$} & \textbf{Valid predictions\newline(total possible)} \\
        \hline
        \hline
            Air-water Temperature Difference & 1.046 & 0.892 & 0.064 & 0.109 & 458687 (488306) \\
            Macro Meteorological & 1.217 & 1.045 & 0.072 & 0.015 & 312367 (488306) \\
            Offshore Macro Meteorological & 0.675 & 0.533 & 0.037 & 0.0 & 317350 (488306) \\
            Persistence & 1.208 & 1.064 & 0.077 & 0.0 & 488306 (488306) \\
            Minute Climatology & 0.625 & 0.502 & 0.036 & 0.021 & 488306 (488306) \\
            Climatology & 0.632 & 0.508 & 0.036 & 0.0 & 488306 (488306) \\
            Hybrid AWT & 0.458 & 0.347 & 0.024 & 0.480 & 458687 (488306) \\
            GBRT & 1.340 & 0.683 & 0.048 & 0.029 & 488306 (488306) \\
            RNN & 0.530 & 0.400 & 0.028 & 0.362 & 488306 (488306) \\
        \hline
    \end{tabular}
    \label{table:9}
\end{table}

As with the \texttt{usna\_cn2\_sm} data set, the minute-level and general climatology models demonstrated similar performance. This may be due to the weaker diurnal cycle described in \cite{jellen2020measurement}. Without hyper-parameter tuning, the random forest model used as a baseline in \cite{jellen2023hybrid} does not outperform the climatology baselines on the validation set. This could be indicative of over-fitting when applied over longer time frames, especially in the context of the model’s performance for the regression task in the \texttt{usna\_cn2\_sm} case. The RNN model's performance is reasonable, but lower than the hybrid air-water temperature difference presented in \cite{jellen2023hybrid}.

\subsection{Forecasting} 

The challenge of forecasting optical turbulence strength differs from regression in that it allows benchmark and future models access to prior measurements of $C_n^2$ when predicting $C_n^2$ at the forecast horizon. As a result, not all regression models are applicable for forecasting tasks. The package currently implements the persistence and climatology baselines, as well as a linear forecasting model based on least-squares regression \cite{numpygenapi2022}, and a model which predicts the mean $C_n^2$ observed within the inference window. These statistical baselines, coupled with the data-driven and deep learning models described in \Cref{sec:baseline_models}.

\subsubsection{\texttt{mlo\_cn2}} 

The MLO $C_n^2$ forecasting task allows benchmark models to observe the most recent 12 observations (1 hour), including measured $C_n^2$, with the objective of predicting $C_n^2$ 6 observations (30 minutes) in the future. The performance of each baseline forecasting model is described in \Cref{table:10}. 

\begin{table}[H]
    \centering
    \caption{Benchmark model performance for the \texttt{mlo\_cn2} forecasting task.}
    \begin{tabular}{|| m{4.5cm} || >{\centering}m{1.2cm} | >{\centering}m{1.2cm} | >{\centering}m{1.2cm} | >{\centering}m{1.5cm} | >{\centering\arraybackslash}m{3.5cm} ||}
        \hline
        \textbf{Model} & \textbf{RMSE} & \textbf{MAE}& \textbf{MAPE} & \textbf{$R^2$} & \textbf{Valid predictions\newline(total possible)} \\
        \hline
        \hline
            Linear & 0.930 & 0.686 & 0.049 & 0.071 & 2432 (2432) \\
            Mean Window & 0.481 & 0.353 & 0.025 & 0.509 & 2432 (2432) \\
            Persistence & 1.227 & 1.089 & 0.077 & 0.0 & 2432 (2432) \\
            Minute Climatology & 0.551 & 0.417 & 0.030 & 0.447 & 2432 (2432) \\
            Climatology & 0.658 & 0.528 & 0.038 & 0.0 & 2432 (2432) \\
            GBRT & 0.428 & 0.318 & 0.023 & 0.652 & 2432 (2432) \\
            RNN & 0.581 & 0.439 & 0.032 & 0.403 & 2432 (2432) \\
        \hline
    \end{tabular}
    \label{table:10}
\end{table}

In the forecasting context, these benchmark models establish a baseline of performance which future models may improve upon. The persistence and climatology models demonstrate lower forecasting skill than the model which predicts the mean $C_n^2$ observed in the inference window. The rapid fluctuation in $C_n^2$ often observed in the boundary layer induces error across all baseline models \cite{sadot1992forecasting} \cite{wang2015prediction}. Despite the long forecast horizon of \si{30 minute} and the \si{15\metre} observation height, these baseline models generate acceptable direct forecasts for $C_n^2$. This performance may, in part, be attributable to the relatively strong diurnal pattern in the MLO $C_n^2$ study \cite{10-26023-CQR2-TQJ9-AH10}. Both the RNN and GBRT models out-performed the statistical baselines, demonstrating reasonable prediction accuracy in direct forecasting. The relative performance improvement was lower for the forecasting task than for the corresponding regression task in \Cref{table:6}. Auto-regressive approaches may further improve these baselines.

\subsubsection{\texttt{usna\_cn2\_sm}} 

The \texttt{usna\_cn2\_sm} data set was used to develop a short-term forecasting task in which the next observation of $C_n^2$ (6 minutes in the future) is predicted using the last 6 observations (36 minutes) during model inference. The diurnal pattern over the Severn River was not as strong as that observed during the MLO $C_n^2$ study, and fluctuations $C_n^2$ were more rapid due to the lower height of observation and other attributes of the propagation environment. The same baseline models used in \Cref{table:10} are evaluated in \Cref{table:11}. 

\begin{table}[H]
    \centering
    \caption{Benchmark model performance for the \texttt{usna\_cn2\_sm} forecasting task.}
    \begin{tabular}{|| m{4.5cm} || >{\centering}m{1.2cm} | >{\centering}m{1.2cm} | >{\centering}m{1.2cm} | >{\centering}m{1.5cm} | >{\centering\arraybackslash}m{3.5cm} ||}
        \hline
        \textbf{Model} & \textbf{RMSE} & \textbf{MAE}& \textbf{MAPE} & \textbf{$R^2$} & \textbf{Valid predictions\newline(total possible)} \\
        \hline
        \hline
            Linear & 0.358 & 0.205 & 0.014 & 0.529 & 4074 (4074) \\
            Mean Window & 0.182 & 0.102 & 0.007 & 0.847 & 4074 (4074) \\
            Persistence & 0.821 & 0.694 & 0.048 & 0.0 & 4074 (4074) \\
            Minute Climatology & 0.453 & 0.362 & 0.026 & 0.120 & 4074 (4074) \\
            Climatology & 0.480 & 0.382 & 0.027 & 0.0 & 4074 (4074) \\
            GBRT & 0.160 & 0.082 & 0.006 & 0.881 & 4074 (4074) \\
            RNN & 0.187 & 0.120 & 0.009 & 0.861 & 4074 (4074) \\
        \hline
    \end{tabular}
    \label{table:11}
\end{table}

As for the MLO $C_n^2$ forecasting task, the model which predicts the mean value of $C_n^2$ observed in the inference window outperforms other persistence and climatology baselines. Due to the short-term nature of forecasts in this task, baseline model error in \Cref{table:11} is lower across all evaluation metrics when compared to the results in \Cref{table:10}. The GBRT model outperforms the mean window model slightly, while performance for the RNN model is comparable to the strongest baseline. The low performance improvement of data-driven and deep learning approaches relative to simpler statistical baselines motivates investigation into architectural improvements for the forecasting context. 

\section{Discussion} 

The \texttt{otbench} package enables robust evaluation and comparison of models for predicting and forecasting optical turbulence strength. The implementation of baseline models allows researchers to compare the performance of new models against statistical, macro-meteorological, data-driven, and deep learning models developed across prior studies \cite{rasp2020weatherbench} \cite{sadot1992forecasting} \cite{wang2015prediction} \cite{jellen2023hybrid}. Differential performance above these benchmarks can also be assessed across more than one field campaign. Researchers can compare their approach across three major measurement campaigns; the package is easily extensible for new data sets and tasks. 

As seen in \Cref{table:6}, \Cref{table:8}, and \Cref{table:9}, data-driven and deep learning models can perform well in predicting the intensity of optical turbulence measured as $C_n^2$ from macro-meteorological data. These results further indicate that baseline models such as the minute-level climatology model perform well, better than the traditional climatology model, in environments with stronger diurnal patterns. This could be further investigated by incorporating a data set and associated regression task captured using a boundary-layer link over a desert environment. In the forecasting context, the model which predicts the mean of $C_n^2$ measurements within the inference window acts as a strong benchmark for the development of future models. This model outperforms both the minute-level climatology model and the general climatology model for both forecasting tasks, as seen in \Cref{table:10} and \Cref{table:11}. The data-driven and deep learning models show small improvements over other baselines, but highlight opportunities for future investigation.

\section{Conclusions} 

Predicting the intensity of optical turbulence, and forecasting it into the future, will enable operational deployment of communication, directed energy, and imaging systems. Despite the importance of these problems, no current benchmarking solution exists for robust evaluation and comparison of regression and forecasting models. To compound this challenge, standard benchmark data sets are not readily available for model development and evaluation. We introduced the \texttt{otbench} package to address these two key challenges. The package's interface for evaluating optical turbulence models on a variety of benchmark tasks and data sets offers researchers the opportunity to understand the performance of their models in context. This includes both the ability to evaluate their models across more than one data set, and to compare performance under a specific task against the prior art. 

The inclusion of persistence, climatology, and other baseline models establishes clearer links with research into novel methods for weather forecasting. \texttt{otbench} is an extensible framework for developing new regression and forecasting models, especially deep learning and machine learning models, while mitigating some risk of over-fitting to specific micro-climates. The data, source code, and package are available at \url{https://github.com/cdjellen/otbench}. 

\section*{Funding} 

Directed Energy Joint Technology Office; Office of Naval Research; Office of Academic Research, United States Naval Academy. 

\section*{Acknowledgments} 

The authors would also like to thank the Waterfront Readiness Center and the United States Naval Academy Mechanical Engineering, Electrical Engineering, Oceanography, and Mathematics Departments for their assistance in conducting the measurement campaign. We are grateful to the NCAR and UCAR data archive teams for making the MLO $C_n^2$ data set available, and to Gordon Maclean and Steve Oncley for their development of the Mauna Loa Seeing Study. The authors would also like to thank the United States Naval Academy Trident Scholar for their support of this research effort.


\bibliographystyle{unsrtnat}

\bibliography{references}

\end{document}